\documentclass{aa}

\usepackage{graphicx}

\usepackage{txfonts}

\usepackage{natbib} 
\bibpunct{(}{)}{;}{a}{}{,}

\usepackage{verbatim} 

\usepackage{hyperref}  

\hypersetup{colorlinks=true,linkcolor=blue,citecolor=blue,filecolor=blue,urlcolor=blue}

\newcommand{\dydz}{\Delta Y/\Delta Z}

\begin{document}

   \title{Constraining the helium-to-metal enrichment ratio $\Delta Y/\Delta Z$ from nearby field stars \\ using {\it Gaia} DR3 photometry}

   \subtitle{}
  \author{N. Ricci \inst{1}, G. Valle \inst{1, 2}, M. Dell'Omodarme \inst{1}, P.G. Prada Moroni
        \inst{1,2}, S. Degl'Innocenti \inst{1,2} 
}
\titlerunning{$\Delta Y /\Delta Z$ form field stars}
\authorrunning{Ricci, N. et al.}

\institute{
        Dipartimento di Fisica "Enrico Fermi'',
        Universit\`a di Pisa, Largo Pontecorvo 3, I-56127, Pisa, Italy
        \and
        INFN,
        Sezione di Pisa, Largo Pontecorvo 3, I-56127, Pisa, Italy
}

   \offprints{G. Valle, valle@df.unipi.it}

   \date{Received 27/07/2025; accepted 12/09/2025}

  \abstract
{}
{We investigate the feasibility of accurately determining the helium-to-metal enrichment ratio, $\dydz$, from {\it Gaia} DR3 photometry for nearby low-mass main sequence field stars.  }
{We selected a sample of 2\,770 nearby MS  stars from the Gaia DR3 catalogue, covering a Gaia $M_G$ absolute magnitude range of 6.0 to 6.8 mag.
We computed a dense grid of isochrones, with $\dydz$ varying from 0.4 to 3.2. These models were then used to fit
the observations using the SCEPtER pipeline.
}
{The fitted values indicated that $\dydz$ values of $1.5 \pm 0.5$ were
adequate for most stars. However, several clues suggested caution ought to be taken in interpreting this result. Chief among these concerns is the trend of decreasing $\dydz$ with increasing $M_G$ magnitude, as well as the discrepancy between the red and blue parts of the observations. This result is further supported by our additional analysis of mock data, which were sampled and fitted from the same isochrone grid. In the mock data, no such trend emerged, while the uncertainty remained as large as 0.7. The robustness of our conclusions  was confirmed by repeating the estimation using isochrones with Gaia magnitudes derived from different atmospheric models and by adopting a different stellar evolution code for stellar model computation.
In both cases, the results changed drastically, clustering at $\dydz \approx 0.4$, which is at the lower end of the allowed values.
}
{Considering the current uncertainties affecting stellar model computations, it appears that
adopting field stars for calibration is not a viable approach, even when adopting precise {\it Gaia} photometry. } 

   \keywords{
Stars: fundamental parameters --
methods: statistical --
stars: evolution --
stars: interiors --
stars: abundances
}

   \maketitle

\section{Introduction}\label{sec:intro}

Although stellar models have significantly improved in terms of predictive accuracy, they are still affected by uncertainties in modelling physical phenomena such as convective transport, microscopic diffusion, and competing processes \citep[see e.g.][]{Viallet2015, Moedas2022}. In addition to these physics-related uncertainties, the assumed chemical composition, particularly the helium abundance ($Y$) and total metallicity ($Z$), significantly impacts stellar model outcomes. Surface metallicity can be accurately determined from absorption lines in stellar spectra; however, direct measurements of helium abundance are generally infeasible for stars cooler than approximately 15,000 K, due to the lack of observable helium spectral lines. Consequently, stellar evolution computations must rely on assumed initial helium abundances. A common approach is to employ a linear relationship between helium abundance and metallicity, $Y = Y_p+\frac{\Delta Y}{\Delta Z} Z$, where $Y_p$ is the primordial helium abundance produced in the Big Bang nucleosynthesis and $\dydz$ is the helium-to-metal enrichment ratio.  

Establishing the precise value of the helium-to-metal enrichment ratio, $\dydz$, has been a subject of extensive investigation. Numerous techniques have been utilised to constrain this parameter, including: making comparisons between theoretical isochrones and observational data in the Hertzsprung-Russell diagram \citep[e.g.][]{pagel98, Casagrande2007, gennaro10, Tognelli2021}; fitting evolutionary tracks to a census of nearby field stars \citep[e.g.][]{jimenez03, Valcarce2013}; using asteroseismic data \citep[e.g.][]{Silva2017, verma2019, nsamba2021}; calibrating from detached, double-lined eclipsing binaries \citep{Ribas2000, Fernandes2012, Valle2024dydz}; developing a standard solar model that accurately replicates the Sun's current luminosity, radius, and surface metallicity-to-hydrogen ratio  \citep[e.g][]{Bahcall1995, Serenelli2010,  Valcarce2012, Vinyoles2017,Buldgen2025}; calibrating stellar models against evolved stars, specifically horizontal branch and red giant stars \citep{Renzini1994, Marino2014, Valcarce2016}; and leveraging the properties of galactic and extragalactic H II regions \citep[e.g.][]{Peimbert1974, Pagel1992, Chiappini1994, Peimbert2000, Fukugita2006, Mendez2020, Kurichin2021} or planetary nebulae \citep{Dodorico1976, Chiappini1994, Peimbert1980, Maciel2001}. The results derived from these diverse methodologies show significant dispersion and are susceptible to systematic biases. For example, standard solar models tend to yield $\dydz$ values below 1, while analyses of evolved stars suggest a range of 2 to 3. Conversely, main sequence (MS) binary stars and comparisons within the HR diagram generally point towards values spanning 1 to 3.

In this paper, we investigate the possibility of constraining $\dydz$ to adopt low-mass  MS stars, profiting from  accurate {\it Gaia} Data Release 3 \citep[DR3;][]{Gaia2021} photometry. Due to their long evolutionary timescale, which makes their position in the colour magnitude diagram almost insensitive to their age, these stars have been already recognised as a valuable target when attempting to carry out such a calibration \citep{jimenez03, Casagrande2007, gennaro10, Valcarce2013}. 
The results are quite variable, ranging from 2 to 5.

The selection of low-mass MS stars limits the sample available for analysis   to nearby stars, because of the faint magnitude of these objects. For our investigation, we selected a sample of nearby stars, within 200 pc, from the {\it Gaia} DR3 catalogue, for which the metallicity [Fe/H] and the $\alpha$-enhancement are known from APOGEE DR17 catalogue \citep{Abdurrouf2022}. The adoption of a single source for chemical abundance data allowed us to avoid possible systematics arising from the use of different catalogues. 
To assess the reliability of the results and to provide a reference of the expected uncertainty owing only to observational errors, we also present a theoretical investigation based on a mock catalogue. This analysis allowed us to establish the minimum variability in the estimated $\dydz$ values in an ideal scenario of a perfect match between stellar models and synthetic objects, apart from the simulated observational uncertainties.

\section{Methods}\label{sec:method}

\subsection{Observational data and selection criteria}\label{sec:data}

Our sample of nearby thin disk stars was selected from {\it Gaia} DR3 data set. Stars with a parallax (corrected according to \citealt{Lindegren2021} zero-points) greater than 5 mas and \texttt{ruwe} lower than 1.4 were selected, resulting in 258\,662 objects.
The sample was cross-matched with the APOGEE DR17 catalogue to extract the metallicity [Fe/H] and $\alpha$- enhancement ${\rm [\alpha/Fe]}$. 
Adopting a conservative approach, stars with ${\rm [\alpha/Fe]}$ greater than 0.05 dex or less than $-0.05$ dex were excluded from the investigation. 
This was justified given the agreement between stellar models and observational data for non-solar $\alpha$-enhancement has been put into question \citep[e.g.][]{Salaris2018, Valle2024age}.
A metallicity [Fe/H] offset of $-0.05$ dex was added to observational data to account for the mismatch between the reference solar mixture adopted in the APOGEE DR17 catalogue ($\log \varepsilon_{\rm Fe} = 7.45 \pm 0.05$) and that included in our investigation ($\log \varepsilon_{\rm Fe} = 7.50 \pm 0.04$). A lower bound on the uncertainty of [Fe/H] was set to 0.06 dex, according to the analysis of systematic differences existing among different surveys \citep[e.g.][]{Hegedus2023, Yu2023}. 
The extinction in the {\it Gaia} DR3 photometric bands was obtained using 
extinction $A_0$ from \citet{Lallement2022} 3D dust maps and extinction coefficients were calculated using the formula presented by \citet{Danielski2018}. 
To balance the need for minimally evolved stars, which reduces the impact of age uncertainties on $\dydz$ estimates, along with the necessity of avoiding significant systematic errors due to the well known discrepancies between observed {\it Gaia} colour-magnitude diagrams and theoretical model isochrones \citep{Brandner2023a, Brandner2023b, Wang2025}, we selected stars within the $M_G$ absolute magnitude range of (6, 6.8) mag. This range mitigates issues arising from the systematic colour deviations observed in the very low-mass regime. 
These discrepancies are not restricted to {\it Gaia} bands and  have been observed by different authors for different photometric systems \citep[e.g.][]{Kucinskas2005,Casagrande2014b}.
The selected sample consisted of 2\,860 objects. 
{Radial-velocity variable stars that remained in the samples, selected according to \citet{Katz2023}, were  removed.\footnote{Radial velocity variable stars were selected based on the criteria 
\texttt{rv\_chisq\_pvalue < 0.01 and rv\_renormalised\_gof > 4 
and rv\_nb\_transits >= 10}. }
Finally, outliers were removed by rejecting stars with {\it Gaia} colour $M_{BP}-M_{RP}$ outside the 5 $\sigma$ range from its mean values computed by dividing the sample in 20 bins in $M_G$ magnitude. The final sample comprises 2\,711 stars. The median metallicity of the sample is [Fe/H] = $-0.04$ dex,  with interquartile interval [$-0.12$, 0.04] dex and range [$-0.58$, 0.35] dex.

\subsection{Stellar model grid}\label{sec:modelli}

\begin{figure}
        \centering
        \resizebox{\hsize}{!}{\includegraphics{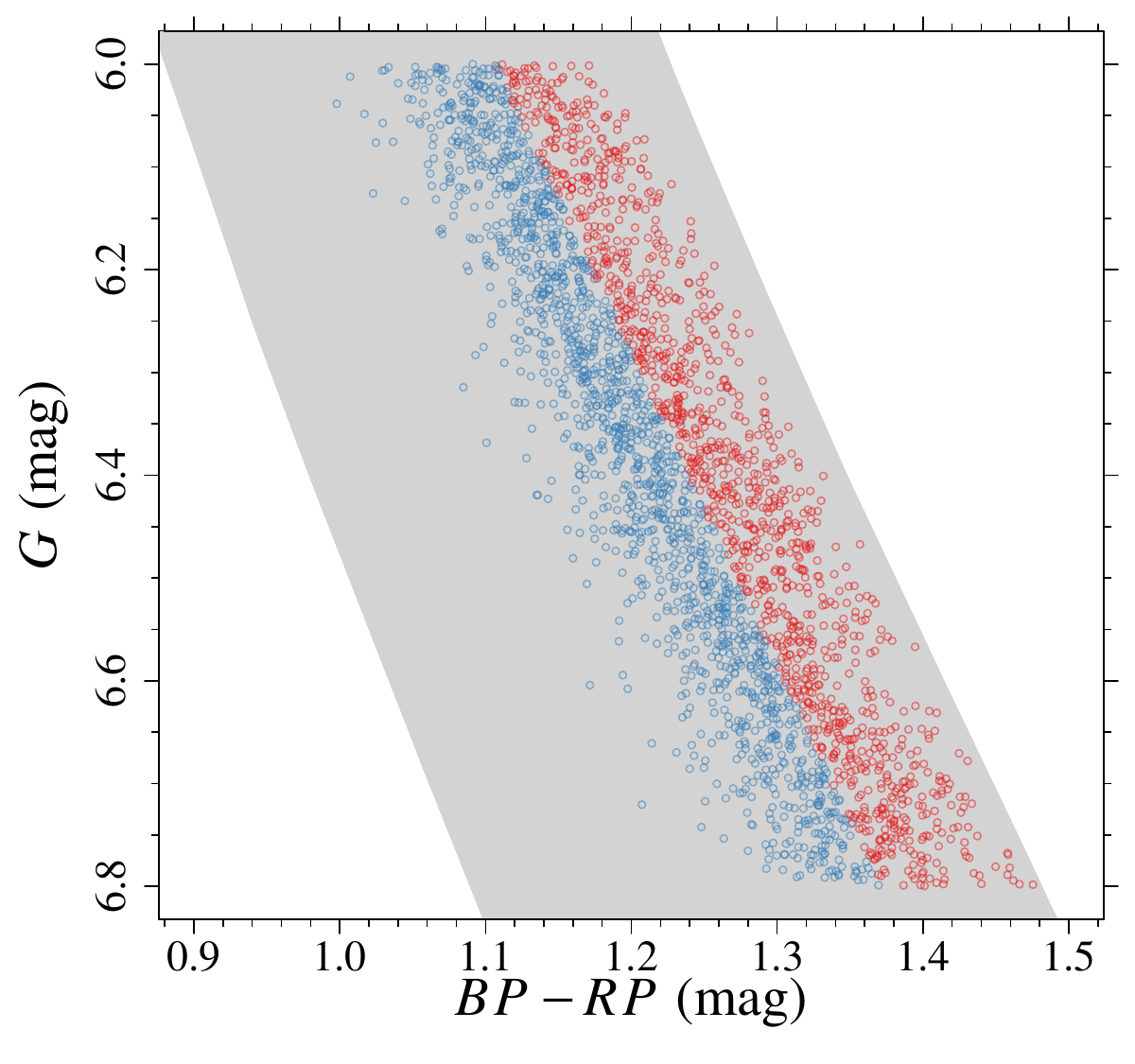}}
        \caption{Colour-magnitude diagram in the $M_G$ vs $M_{BP}-M_{RP}$ plane of the selected stars. Different colours indicate the red and the blue part of the sample (see text). The shaded grey region indicates the area encompassed by the computed stellar isochrones.  }
        \label{fig:cmd}
\end{figure}

The grid of stellar evolutionary models was calculated for the 0.50 to 1.00 $M_{\sun}$ mass range with a step of 0.025 $M_{\sun}$, spanning the evolutionary stages from the pre-main sequence to the onset of the red giant branch (RGB).
The initial metallicity [Fe/H] ranged from $-0.5$ dex to 0.35 dex with
a step of 0.025 dex. 
We adopted the solar heavy-element mixture by \citet{AGSS09}. 
For each metallicity, we considered a range of initial helium abundances based on the commonly used linear relation, $Y = Y_p+\frac{\Delta Y}{\Delta Z} Z$,
with the primordial helium abundance,  $Y_p = 0.2471 \pm 0.001$, from \citet{Planck2020}.
The helium-to-metal enrichment ratio, $\Delta Y/\Delta Z$, was varied
from 0.4 to 3.2 with a step of 0.1. 

The models were computed with the FRANEC code, in the same
configuration previously adopted to compute the Pisa Stellar
Evolution Data Base\footnote{\url{http://astro.df.unipi.it/stellar-models/}} 
for low-mass stars \citep{database2012}. The outer boundary conditions  were established by the solar semi-empirical $T(\tau)$  of  \citet{Vernazza1981}, which aptly approximate the results obtained using the hydro-calibrated $T(\tau)$  \citep{Salaris2015, Salaris2018}. 
The models were computed
assuming the solar-scaled mixing-length parameter $\alpha_{\rm
        ml} = 2.02$.
Atomic diffusion was included by adopting the coefficients given by
\citet{thoul94} for gravitational settling and thermal diffusion. 
To prevent extreme variations in surface chemical abundances, 
the diffusion velocities were
multiplied by a suppression parabolic factor that is equivalent to one for 99\% of the mass of the structure and zero at the base of the atmosphere \citep{Chaboyer2001}.

The raw stellar evolutionary tracks were reduced to a set of isochrones spanning the age range from 0.1 Gyr to 11.5 Gyr, the estimated age range of the Galactic disk according to recent literature analyses \citep[e.g.][]{fantin2019, gallart2024}.
Bolometric corrections used to derive the {\it Gaia} magnitudes were obtained from the PHOENIX2011 grid \citep{Allard2011}, which encompasses the effective temperatures from $400\,\mathrm{K}$ to $100\,000\,\mathrm{K}$ and surface gravities between $-0.5$ and $5.5$ dex.

\subsection{Fitting technique}\label{sec:fit-method}

The analysis was carried out using the SCEPtER pipeline\footnote{Publicly available on CRAN: \url{http://CRAN.R-project.org/package=SCEPtER}.}, a well-tested technique to fit single and binary systems
\citep[e.g.][]{scepter1,eta}. 
The pipeline estimates  the parameters of interest (i.e. the stellar age and its initial chemical abundances), adopting a grid maximum-likelihood  approach.

Here, we provide only a sketch of the method for convenience. For every $j$-th point in the fitting grid of precomputed stellar models, a likelihood estimate is obtained as
\begin{equation}
        {{\cal L}}_j = \left( \prod_{i=1}^n \frac{1}{\sqrt{2 \pi}
                \sigma_i} \right) 
        \times \exp \left( -\frac{\chi^2}{2} \right)
        \label{eq:lik}
        ,\end{equation}
\begin{equation}
        \chi^2 = \sum_{i=1}^n \left( \frac{o_i -
                g_i^j}{\sigma_i} \right)^2
        \label{eq:chi2},
\end{equation}
where $o_i$ are the $n$ observational constraints, $g_i^j$ are the $j$-th grid point corresponding values, while $\sigma_i$ represents the observational uncertainties. Absolute {\it Gaia} magnitudes and metallicity [Fe/H] were used as constraints.

For each star, we adopted the maximum a posteriori (MAP) estimates of stellar age and $\dydz$ as the fit values. The MAP estimate corresponds to the mode of posterior density, which is equivalent to the mode of the likelihood function when no prior information is incorporated.
To mitigate the effects of grid discretisation, we calculated the estimates by averaging the respective quantities from all models with likelihoods exceeding $0.975
\times {\cal L}_{\rm max}$\footnote{We verified that the choice of a threshold value is not critical, as any value above 0.95 produces negligible differences. }. 
We selected the MAP estimate over the mean or median of the posterior density due to the significant variance observed in our estimates. In many cases, the posterior densities were moderately peaked and did not approach zero at the limits of the allowed $\dydz$ range. Consequently, using the mean or median would introduce a strong dependence on the specified parameter range, a problem that was avoided by using the MAP estimate.

\section{Estimated $\dydz$}\label{sec:results}

\begin{figure}
        \centering
        \resizebox{\hsize}{!}{\includegraphics{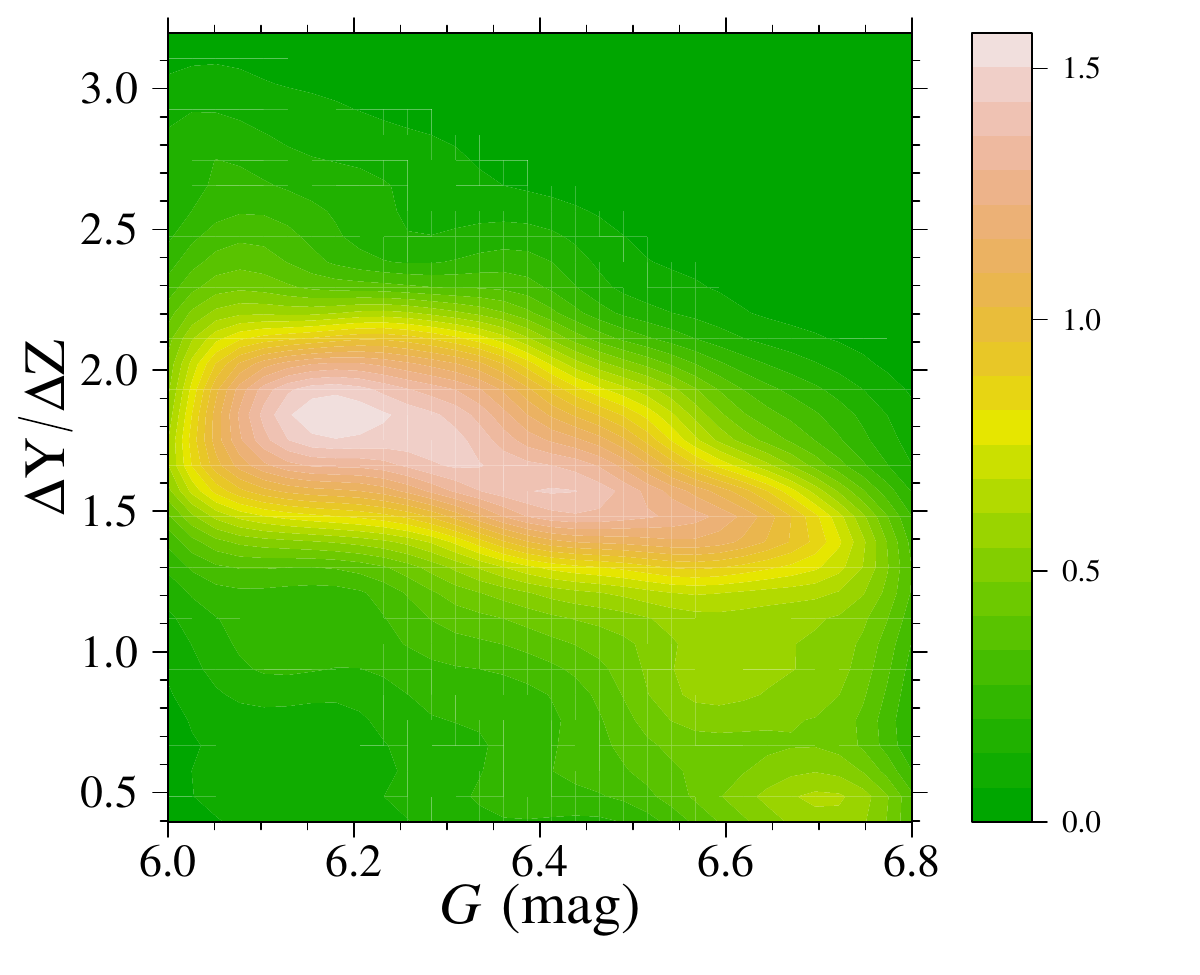}}\\
    \resizebox{\hsize}{!}{\includegraphics[width=8.2cm]{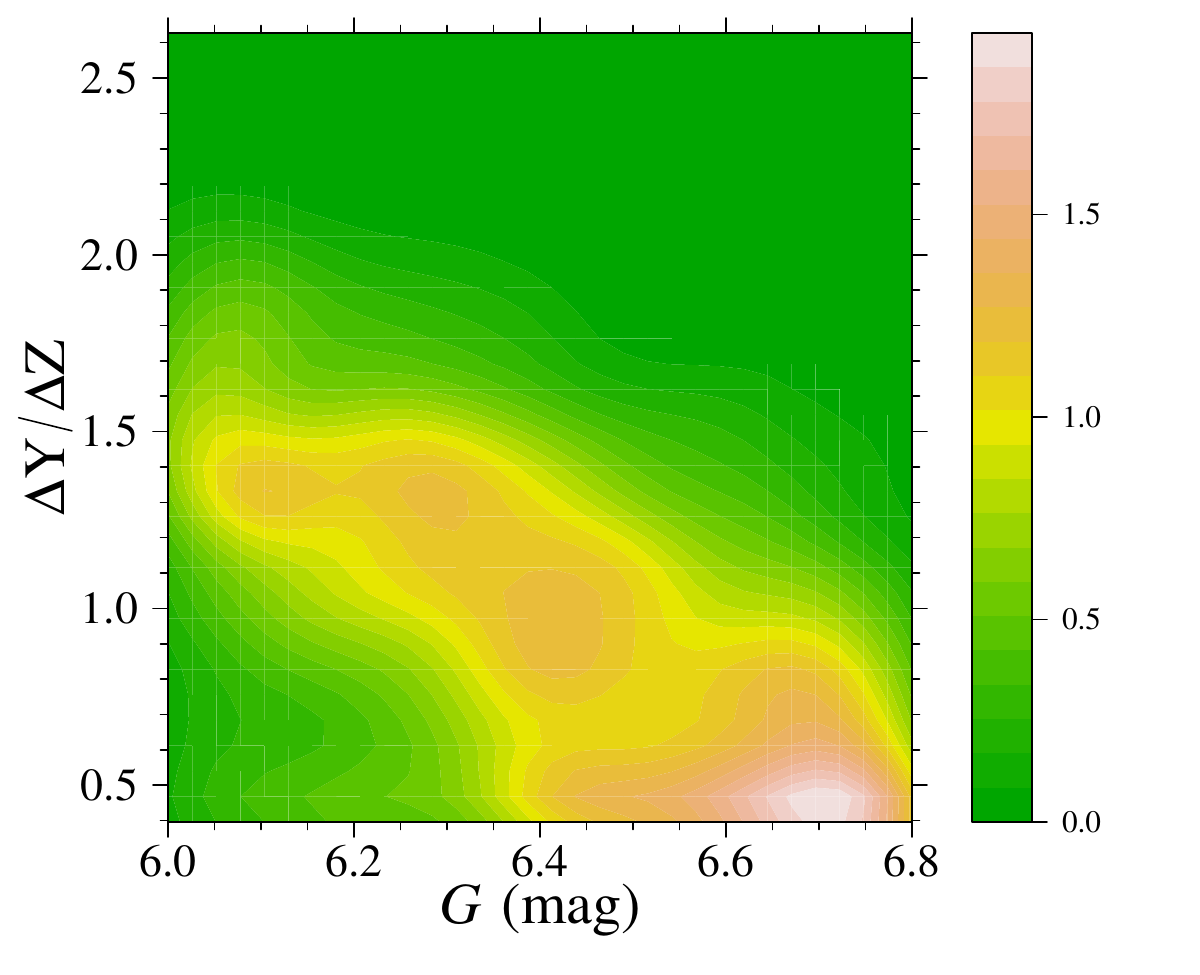}}
        \caption{Bidimensional posterior density estimates in the $\dydz$ versus $M_G$ plane. {\it Top}: Density of the blue sub-sample. {\it Bottom}: Density of the red component.   }
        \label{fig:d2}
\end{figure}

\begin{figure}
        \centering
        \resizebox{\hsize}{!}{\includegraphics{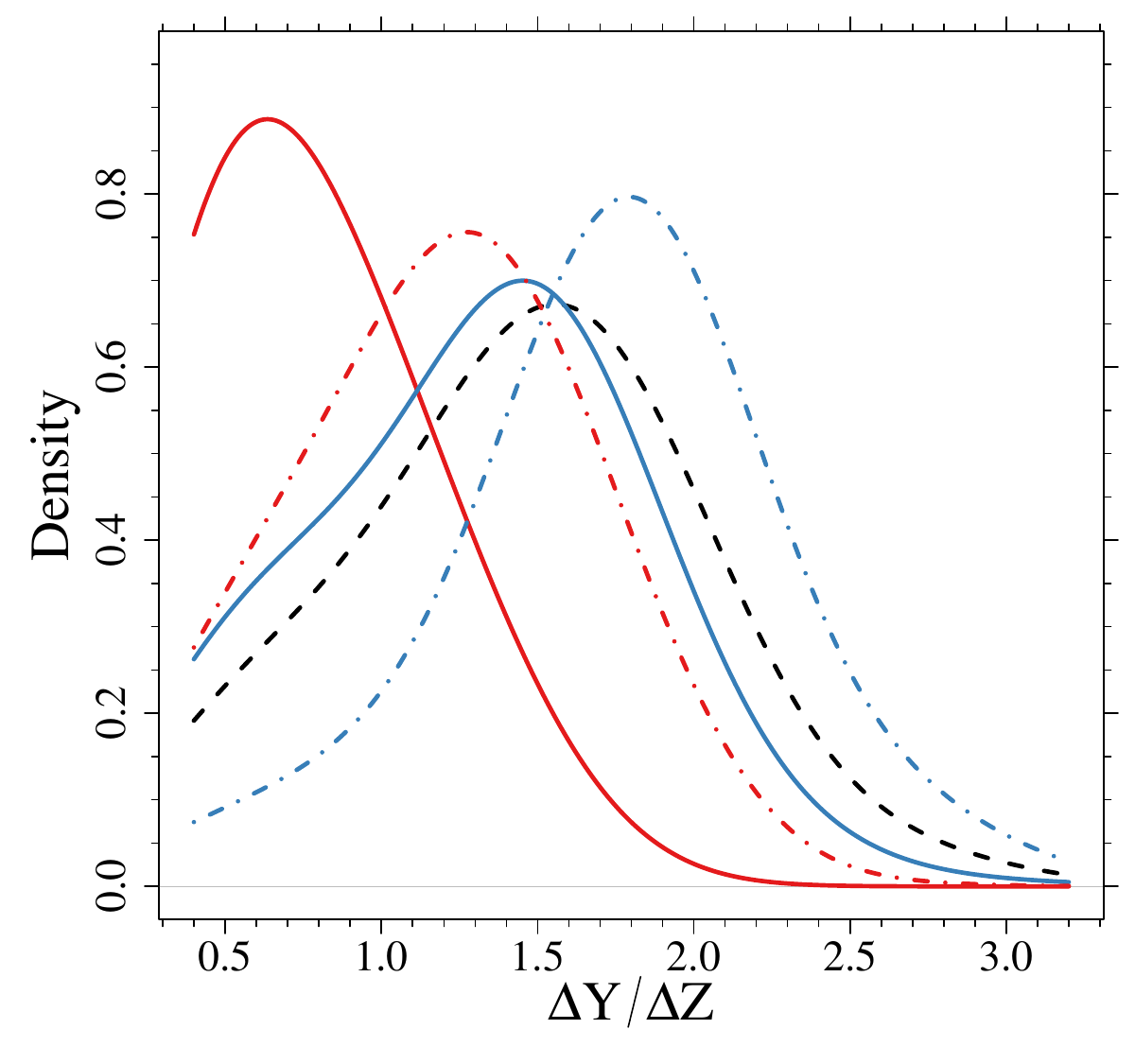}}
        \caption{Kernel density estimators for the analysed sub-samples. The solid red and blue lines indicate the red and blue sub-samples for $M_G > 6.4$ mag. The dot-dashed lines corresponds to the red and blue sub-sample but for $M_G < 6.4$ mag. The black dashed line is the global estimator for all the stars but for the less evolved red components.   }
        \label{fig:d1}
\end{figure}

The selected sample was divided into two parts, according to colour $M_{BP}-M_{RP}$. The median colours in the range of (6.00, 6.05) mag and (6.75, 6.80) mag were calculated and the points $P_1 =(6.00, 1.11)$ mag, $P_2 =(6.80, 1.37)$ mag were identified. The stars were then classified as blue (red) component if their $M_G$ magnitude was higher (lower) than that resulting from the straight line passing through $P_1$ and $P_2$.
The sub-samples are shown in Fig.~\ref{fig:cmd}. From this figure, it is apparent to the  naked eye that the data approach the red edge of the computed isochrones as $M_G$ increases, offering evidence that  the systematic distortion discussed in Sect.~\ref{sec:data} begins to manifest itself in this range.

The fit revealed a relevant difference between the two sample sub-populations. Although the $\dydz$ values estimated from the bluer component are generally well placed within the grid, this is not true for the redder component, especially for the $M_G \geq 6.4$ mag. In this case, the results are generally located at the lower edge of the grid. This behaviour is evidenced in Fig.~\ref{fig:d2}, where the bidimensional estimators of the posterior probability in the $\dydz$ versus $M_G$ plane are shown. It is apparent that for the bluer component, the fit generally suggests $\dydz$ values between 1.5 and 2.0, with a slight tendency to obtain lower values at high $M_G$. In contrast, the corresponding plot for the redder component shows a strong systematic decrease, which affects $\dydz$ estimates for $M_G$ greater than about 6.4 with a clustering at the lower edge of the grid. Moreover, even for $M_G < 6.2$ mag, where the results are almost stationary and independent of $M_G$, the estimated $\dydz$ is lower than the values inferred from the bluer component by about 0.6.

Figure~\ref{fig:d1} shows the kernel density estimators for the posterior in four groups, obtained by classifying stars in blue and red components and low and high $M_G$ magnitude, adopting $M_G = 6.4$ mag as cut-off value. The anomalous inference from the redder and low luminosity stars is apparent. By excluding these stars from the sample, we got an estimate of $\dydz = 1.44 \pm 0.54$, while the estimate over the blue component alone is $\dydz = 1.58 \pm 0.54$.

\section{Analysis of the result robustness}\label{sec:mesa}

The preceding analysis casts significant doubt on the reliability of the derived results. Although the blue component and (to a lesser extent) the more evolved red component hint at a $\dydz \approx 1.5$ value as a potential descriptor for the sample, several inconsistencies warrant further investigation. Firstly, a distinct trend emerges for lower luminosity stars within both the red and blue components. In these instances, less evolved stars consistently suggest a $\dydz$ value lower than their more evolved counterparts. 
Secondly, the median $\dydz$ values obtained from the blue and red components exhibit a statistically significant difference (Wilcoxon test, $P<10^{-16}$). This result holds even considering only more evolved stars, with $M_G < 6.2$ mag. This disparity raises concerns about the consistency of the results. This section explores the possibility that these discrepancies are either inherent to the methodology or stem from systematic discrepancies between the stellar model grid and the observational data.

\begin{figure}
        \centering
        \resizebox{\hsize}{!}{\includegraphics{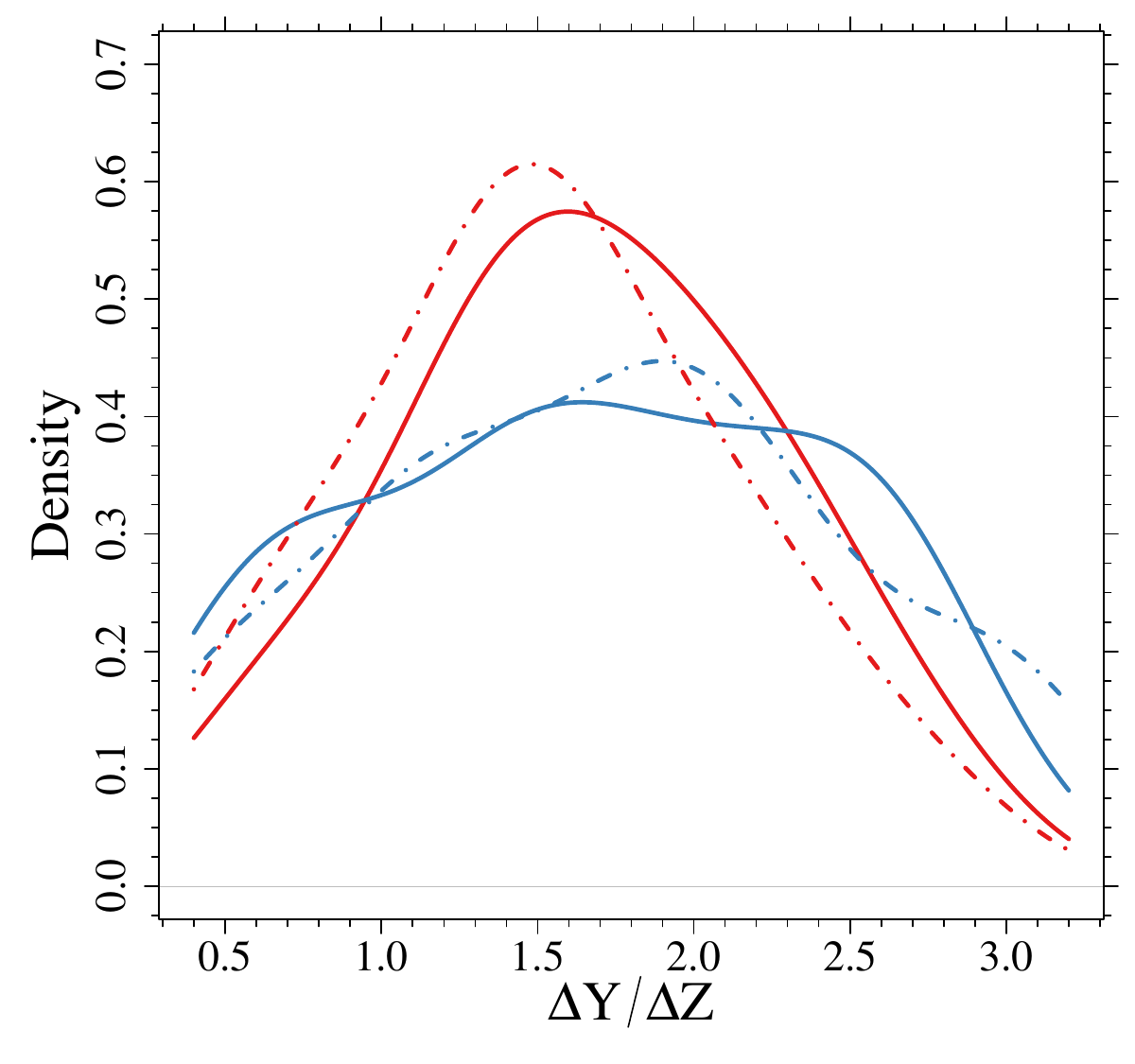}}
        \caption{Same as in Fig.~\ref{fig:d1}, but for mock stars sampled from the grid with $\dydz = 1.8$.  }
        \label{fig:d1mock}
\end{figure}

\subsection{Mock data analysis}

To examine the expected results under ideal conditions, where observations and model predictions align except for random discrepancies attributable to observational uncertainties, we replicated the estimation procedure outlined in Sect.~\ref{sec:results}. We utilised a mock sample, comprising 400 objects randomly drawn from the stellar model grid, spanning the same $M_G$ and $M_{BP}-M_{RP}$ ranges as the observed stars. All objects were sampled from the grid with $\dydz=  1.8$. Subsequently, Gaussian random perturbations to the {\it Gaia} magnitudes were added, simulating observational errors. These perturbations adopted standard deviations corresponding to the median uncertainties of the nearby star sample ($\sigma_{M_G} = 0.0052$ mag, $\sigma_{M_{BP}} = 0.0054$ mag, $\sigma_{M_{RP}} = 0.0053$ mag).

The results of this mock data analysis are presented in Fig.~\ref{fig:d1mock}. The analysis demonstrates the viability of the methodology, as the estimated $\dydz$ values closely approximate the true value of 1.8. However, it also reveals that even under ideal conditions, the results are subject to significant uncertainty, solely due to observational errors.  The overall mean estimated helium-to-metal enrichment ratio was $\dydz = 1.66 \pm 0.69$. A slight underestimation bias of $-0.14$ is observed, representing approximately 20\% of the random variability. This tendency, previously noted in analyses of 
different objects and evolutionary phases \citep[e.g.][]{bulge,cluster2018}, arises from the unequal displacement in the {\it Gaia} magnitude space between homologous isochrones with different $\dydz$. In fact, all other inputs being equal, a symmetric variation of $\dydz$ does not result in a symmetric displacement in the ($M_G$, $M_{BP}$, $M_{RP}$) space. In particular, there is an increasing separation between homologous isochrones when $\dydz$ increases. This is the underlying asymmetry that results in the observed bias. Consistently with the analysis presented in Sect.~\ref{sec:results}, the estimate of $\dydz$ for the blue component, $\dydz = 1.71$, is higher than that of the red sub-sample, $\dydz = 1.59$. However, this difference is considerably smaller than the 0.6 disparity observed for the evolved stars in the real sample. Crucially, no trend with $M_G$ magnitude is evident in the mock data analysis. This finding supports the conclusion that the trend observed in Section \ref{sec:results} is spurious, stemming from the systematic displacement between isochrones and data, where the former appear bluer at higher magnitudes.

\subsection{Estimates adopting different atmosphere models}\label{sec:marcs}

\begin{figure}
        \centering
    \resizebox{\hsize}{!}{\includegraphics{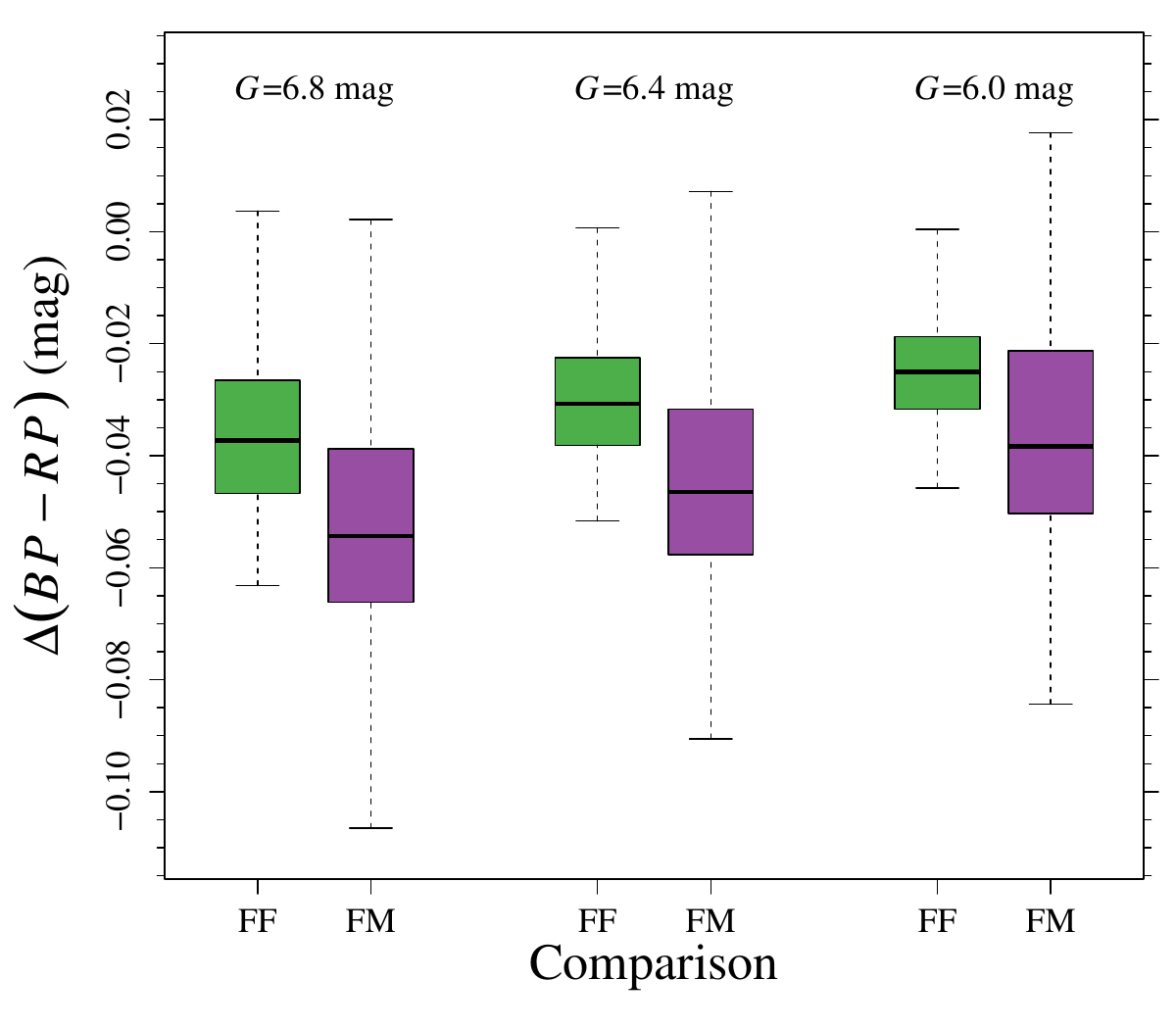}}
        \caption{Boxplot of the differences in the colour index between 
    FRANEC isochrones computed with MARCS2008 models and the reference PHOENIX2011 grid (comparison FF, green boxes), and between MESA and FRANEC reference isochrones (comparison FM, violet boxes) at three different levels of $M_G$ magnitude. }
        \label{fig:box-mesa}
\end{figure}

\begin{figure}
        \centering
        \resizebox{\hsize}{!}{\includegraphics{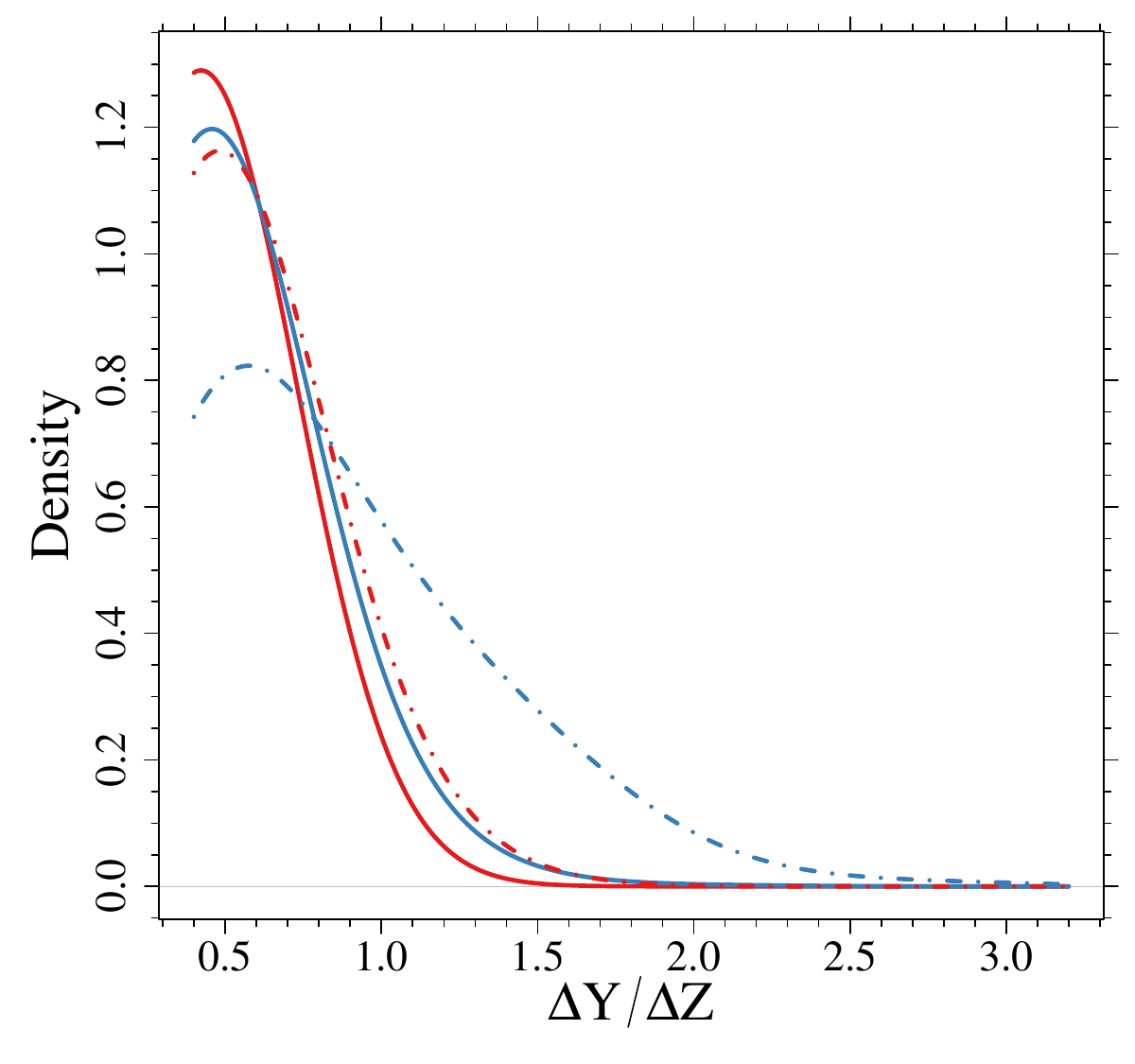}}
        \caption{Same as in Fig.~\ref{fig:d1}, but adopting MARCS2008 models to obtain isochrone {\it Gaia} magnitudes.  }
        \label{fig:d1marcs}
\end{figure}

The accuracy of the $\dydz$ estimates derived from field stars in Sect.~\ref{sec:results} is primarily determined by the consistency between the {\it Gaia} band observations and the stellar model grid. However, non-negligible variability persists in stellar model computations due to the freedom stellar modellers have in adopting different input physics within their allowed uncertainties \citep[see e.g.][]{incertezze1, Stancliffe2016}. For determining $\dydz$ from unevolved stars, the selection of bolometric correction tables is of paramount importance \citep{Plez2011J},
because the choice of atmospheric models is  among the most important sources of variability when comparing isochrones and data \citep[see e.g.][]{goodness2021}.
The results presented in Sect.~\ref{sec:results} were determined on the basis of the atmosphere models of \citet{Allard2011}.  We  repeated the estimation of $\dydz$ using the same stellar model grid, but with {\it Gaia} magnitudes computed from the MARCS synthetic spectra \citep{Gustafsson08}, as described below.

The MARCS2008 isochrones exhibited a systematic offset in $M_{BP}-M_{RP}$ colour  compared to the PHOENIX2011-based models. We defined $\Delta(M_{BP}-M_{RP})$ as the difference in the $M_{BP}-M_{RP}$ colour index between MARCS2008 and PHOENIX2011 isochrones, where a negative value indicates that the MARCS2008 isochrones have a lower colour index. 
Figure~\ref{fig:box-mesa} illustrates $\Delta(M_{BP}-M_{RP})$ calculated at three distinct $M_G$ magnitudes: 6.0, 6.4, and 6.8 mag. The comparisons are identified by the label: FF.
The median $\Delta(M_{BP}-M_{RP})$ was approximately $-0.037$ mag for less evolved stars and $-0.025$ mag at $M_G=6.0$ mag. This substantial displacement, exceeding the observational uncertainty by a factor greater than five, significantly impacted the fitting process. Specifically, apart from more evolved blue stars, all stars were fitted near the lower $\dydz$ boundary of the grid, as shown in Fig.~\ref{fig:d1marcs}.
This result suggests that the choice of atmosphere models plays a dominant role in the fit and ultimately cautions against attempting to calibrate $\dydz$ for unevolved MS stars.

\subsection{Estimates adopting MESA isochrones}

\begin{figure}
        \centering
        \resizebox{\hsize}{!}{\includegraphics{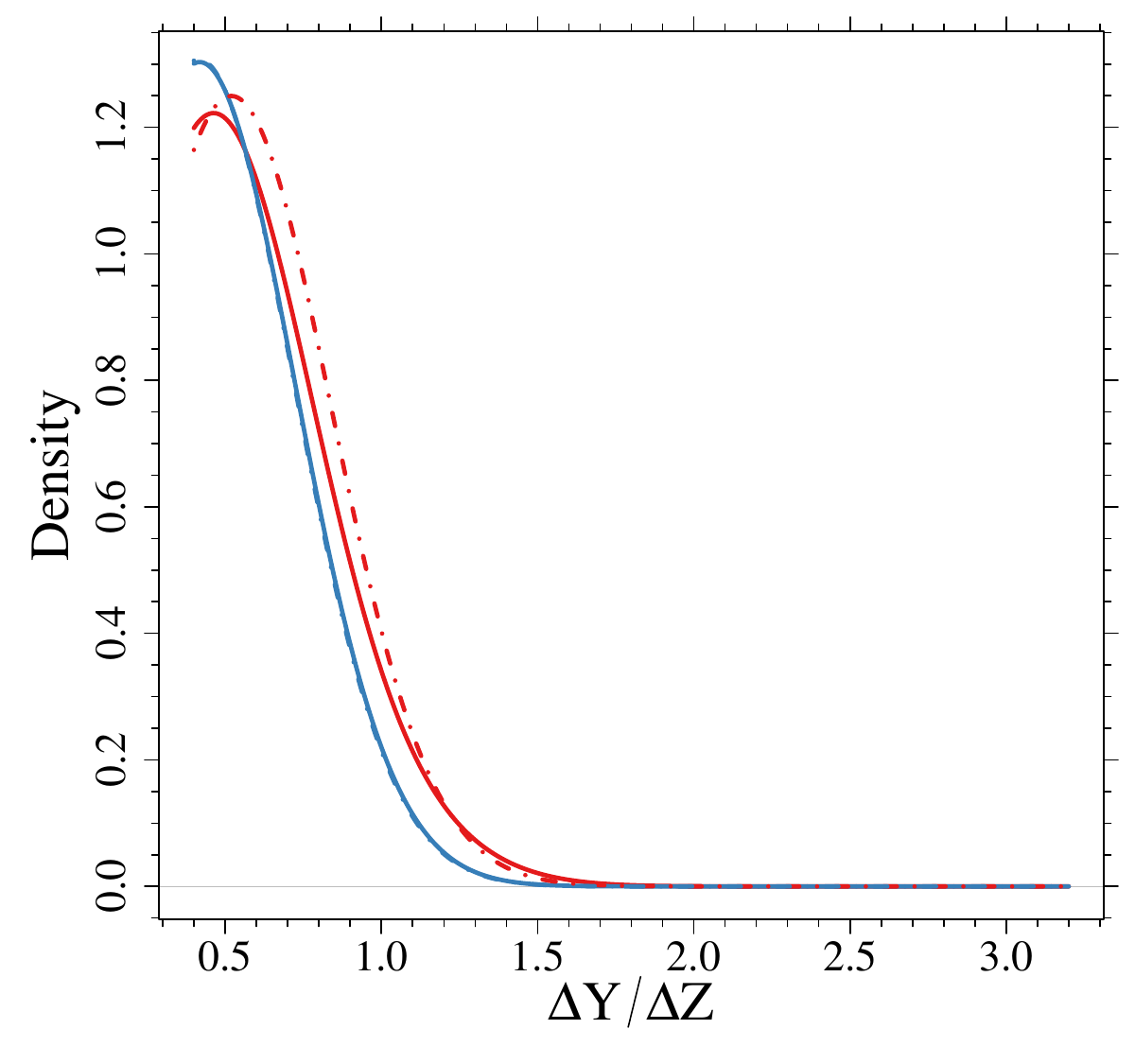}}
        \caption{Same as in Fig.~\ref{fig:d1}, but adopting MESA isochrones for the fit.  }
        \label{fig:d1mesa}
\end{figure}

As a final validation, we repeated the $\dydz$ estimate using an alternative stellar evolutionary code and bolometric correction tables. Specifically, we used MESA release r24.03.01 \citep{MESA2018, Jermyn2023} to compute isochrones, mirroring the setup in Sect.~\ref{sec:modelli}. This included adopting the \citet{AGSS09} solar mixture, modelling stars from 0.5 to 1.0 $M_{\sun}$ with microscopic diffusion, and using a solar-calibrated mixing length parameter $\alpha_{\rm ml} = 2.32$\footnote{For the calibration the metallicity, $Z$, initial helium abundance, $Y$, and mixing-length parameter were iteratively refined using a Levenberg-Marquardt algorithm, aiming to reproduce the observed solar radius, luminosity, logarithm of the effective 
temperature, and surface $Z/X$ at the Sun's current age.}. The boundary conditions were established using the solar-calibrated Hopf $T(\tau)$ relation \citep{MESA2013}, derived from solar atmosphere model C from \citet{Vernazza1981}. Isochrones were computed using the \texttt{iso} code\footnote{\url{https://github.com/aarondotter/iso}.} \citep{Dotter2016} accessible via the MIST website.
For bolometric corrections, we utilised tables from the MIST web page\footnote{\url{https://waps.cfa.harvard.edu/MIST/model\_grids.html\#bolometric}.}, computed from the C3K grid \citet{Conroy2018} of 1D atmosphere models based on ATLAS12/SYNTHE \citep{Kurucz1970S, Kurucz1993}.

Consistentkt with the findings of the previous section, MESA isochrones exhibited a systematically lower $M_{BP}-M_{RP}$ colour compared to the reference FRANEC  models (see comparison with FM in Fig.~\ref{fig:box-mesa}).
The median $\Delta(M_{BP}-M_{RP})$ was approximately $-0.055$ mag for less evolved stars and $-0.038$ mag at $M_G=6.0$ mag. Due to these large differences, nearly all stars were fitted near the lower $\dydz$ boundary of the grid, as shown in Fig.~\ref{fig:d1mesa}.

\section{Conclusions}\label{sec:conclusions}

In this work, we tried to constrain the helium-to-metal enrichment ratio, $\dydz$, using low-mass MS stars. This technique has previously been employed in the literature \citep[e.g.][]{jimenez03, gennaro10, Valcarce2013} and yielded a wide range of plausible values, from 2 to 5.
We explored the possibility of narrowing this allowed range by leveraging  precise {\it Gaia} photometry and  metallicity data from APOGEE DR17 \citep{Abdurrouf2022} as the observational constraints.

To this aim, we selected a sample of 2\,711 nearby low-mass MS stars from the {\it Gaia} DR3 catalogue \citep{Gaia2021}, spanning a {\it Gaia} $M_G$ absolute magnitude range of 6.0 to 6.8 mag. We computed a dense grid of approximately 12\,000 isochrones, with $\dydz$ varying from 0.4 to 3.2 and metallicity [Fe/H] ranging from $-0.5$ to 0.35 dex.
These models were then used to fit the observations using the SCEPtER pipeline \citep{scepter1,eta}.

The fitted values indicated that a $\dydz$ of $1.5 \pm 0.5$ was adequate for the majority of the stars, with the notable exception of the less evolved and redder stars. However, several clues suggested caution in interpreting this result. The most important of these were the trend of decreasing $\dydz$ with increasing $M_G$ magnitude and the discrepancy in $\dydz$ derived from the red and blue parts of the observations. These issues are most likely caused by the well-known discrepancies between synthetic colours and observations for less evolved stars, which affect {\it Gaia} and other photometric systems \citep{Casagrande2014b, Brandner2023a, Brandner2023b, Wang2025}. Supporting evidence comes from the analysis of mock data, which were sampled and fitted from the same isochrone grid. In this case, no trend emerged and the $\dydz$ value was consistently estimated to match the values adopted in the sampling procedure, with a slight bias of $-0.14$ and a significant 0.7 variability due to the simulated observational uncertainties.

To further investigate the robustness of the results, we performed two additional tests. First, we repeated the estimation using isochrone with {\it Gaia} magnitudes derived from different atmosphere models. Specifically, we substituted the reference PHOENIX2011 models \citep{Allard2011} with the MARCS2008 models \citep{Gustafsson08}. Given that the MARCS2008-based grid consistently yields a lower $M_{BP}-M_{RP}$ colour index, the fitted $\dydz$ changed drastically, clustering at 0.4, the lower end of the allowed values.
Second, we obtained a similar result by adopting a different stellar evolution code that is, MESA r24.03.01 \citep{MESA2018, Jermyn2023} coupled with C3K atmosphere models \citep{Conroy2018}, obtained from the MIST Web page.

The strong dependence of the fitted $\dydz$ on the adopted stellar models and bolometric correction tables suggests caution in interpreting the obtained results. Given the current uncertainties affecting stellar model computations \citep[e.g.][]{incertezze1, Stancliffe2016} and the significant discrepancies among available sets of bolometric correction tables, it appears that adopting field stars for calibration is not viable. This conclusion remains unchanged even when considering recent efforts to calibrate empirical colours in the {\it Gaia} bands using clusters as our references \citep[e.g.][]{Wang2025}. Although this calibration reconciles observed and predicted colour indices, isochrones calibrated in this way are not suitable for estimating other parameters such as $\dydz$. Since  this colour calibration inherently assumes the specific input physics and chemical patterns of the models used in the calibration, it would obviously be biased toward these values.

\begin{acknowledgements}
G.V., P.G.P.M. and S.D. acknowledge INFN (Iniziativa specifica TAsP) and support from PRIN MIUR2022 Progetto "CHRONOS" (PI: S. Cassisi) finanziato dall'Unione Europea - Next Generation EU.
\end{acknowledgements}

\bibliographystyle{aa}
\bibliography{biblio}

\end{document}